# Electrically Induced Photonic and Acoustic Quantum Effect From Liquid Metal Droplets in Aqueous Solution

Qian Wang,[1] Yang Yu[2] and Jing Liu[1,2]*

[1]Technical Institute of Physics and Chemistry, Chinese Academy of Sciences,
Beijing 100190, China.

[2]Department of Biomedical Engineering, School of Medicine, Tsinghua University,
Beijing 100084, China.

*Corresponding author. E-mail: jliu@mail.ipc.ac.cn.

**Abstract:**

Basically, any physical properties of a matter can find root in its quantum mechanics. So far, several macroscopic quantum phenomena have been discovered in the Josephson junction. Through introducing such a structure with a liquid membrane sandwiched between two liquid metal electrodes, we had ever observed a lighting and sound phenomenon which was explained before as discharge plasma. In fact, such an effect also belongs to a quantum process. It is based on this conceiving, we proposed here that an electrically controllable method can thus be established to generate and manipulate as much photonic quantum as desired. We attributed such electrically induced lighting among liquid metal droplets immersed inside aqueous solution as photonic quantum effect. Our experiments clarified that a small electrical voltage would be strong enough to trigger blue-violet light and sound inside the aqueous solution system. Meanwhile, thermal heat is released, and chemical reaction occurs over the solution. From an alternative way which differs from former effort in interpreting such effect as discharge plasma, we treated this process as a quantum one and derived new conceptual equations to theoretically quantify this phenomenon in light of quantum mechanics principle. It can be anticipated that given specific designing, such spontaneously generated tremendous quantum can be manipulated to entangle together which would possibly help mold functional elements for developing future quantum computing or communication system. With superior adaptability than that of the conventional rigid junction, the present electro-photonic quantum generation system made of liquid metal droplets structure could work in solution, room temperature situation and is easy to be adjusted. It suggests a macroscopic way to innovate the classical strategies and technologies in generating quantum as frequently adopted in classical quantum engineering area.

## 1. Introduction

The concept of quantum was first proposed by Max Planck in 1900. The fundamental notion of this theory lies in that the energy can be treated as a discrete value consisting of integer multiples of one quantum. With further endeavors from world-wide researchers, many physical properties have been gradually disclosed as discontinuous quantization, such as photon. In 1905, Einstein introduced the concept of light quantum and explained the photoelectric effect, which is well-known that when light hits the surface of a metal, the electrons inside the metal are able to be ejected through absorption of light energy. [1,2] Nowadays, more and more researchers have accepted this theory. In short, all tangible physical properties could be quantized when explaining in quantum terms.

The recently emerging quantum computer is a device that performs operation, storage and processing of quantum information via the law of quantum mechanics. Its core component is generally a superconducting electronic device called Josephson junction, which is composed of a superconductor-insulator-superconductor structure. [3,4] The insulator usually formed by metal oxides is rather thin (<10nm), which plays an important role on the generation of Josephson tunneling current. So far, lots of macroscopic quantization phenomena including energy levels quantization and quantum tunneling have been observed in the circuits with Josephson junction. [5-9] However, until now, nearly all the existing junctions are rigid. Thus, the thickness of middle insulator layer is not easy to be adjusted once formed.

Liquid metal is an unconventional soft material with both high conductivity of the metal and flexibility and deformability of the fluid at room temperature. Nowadays, investigations on the liquid metal fundamentals and applications have become scientific hot topics and frontiers over the world. These liquid metal functional materials could offer extraordinary properties in various application fields, such as energy, thermal management, advanced manufacture, biomedicine, and so on. [10-15] Since it is mobile, the surface of liquid metal could achieve a perfect smoothness of atomic level. And considering the thin oxide layer spontaneously formed on the surface, the low melting liquid metal could be adopted as a potential material for junction structure used in quantum computer. [16] The present group has proposed a transformable structure of liquid membrane sandwiched between the two liquid metal electrodes. Different from the classical Josephson junction, such kind of liquid metal sandwich structure [16] or an extended droplet circuit system [17] could be adjusted under external factors. Depending on the applied electric field, the thickness of liquid interface membrane could reach a minimum scale or even disappear completely, [18] and the conductivity of both liquid metal electrodes will also correspond with the change of the structure. Hereby, we had ever put forward a hypothesis that if the thickness of middle liquid film is controlled within a certain range, liquid quantum tunneling effect will be expected to appear. [16] A conceptually innovative all-soft or liquid quantum device which breaks through the traditional rigid quantum device was proposed. Not only that, some fabrication methods were also suggested to make liquid quantum device. However, no quantum tunneling effect was directly measured there. Following that, Ren et al. proposed the liquid metal enabled droplet circuits based on such all-soft structure, which could achieve the volumetric conductive functions throughout the whole environment. [17]

Recently, we found a phenomenon which appears like discharge plasma. [19] It is easily triggered in aqueous solution under a low voltage via the liquid metal electrode that is either static or a jetting stream. Meanwhile, blue, violet and ultraviolet light was generated, and thermal heat was released. It was noticed that such phenomenon often occurs at the gap of liquid metal stream, which could also be generated by using the sandwich structure. In fact, just like the classical photoelectric effect, the electrically induced lighting from the liquid metal droplet should also belong to a quantum process. It is for this reason, we are dedicated here to propose to adopt such liquid metal system to generate and manipulate quantum. Along with that, we then put forward a theory to quantify and interpret this phenomenon via light quantum terms. Its potential prospect in developing future quantum computing or communication system was also discussed.

## 2. Electro-Photonic Quantum Effect Induced From Liquid Metal System

Figure 1 presents some violet lights that are created using liquid metal under different conditions at room temperature. The liquid metal ($Ga_{75.5}In_{24.5}$ alloy) here serves as a soft electrode and is connected with the positive pole at least. The other end connected with the negative pole could be rigid or soft metal electrode. As shown in Figure 1a, a copper plate is used for negative electrode, and the whole process is carried out under the water. The liquid metal electrode hardly contacts the copper plate when there is a light. For the case in Figure 1b and Figure 1c, the negative electrode is also the copper but the liquid metal electrode is ejected from the positive towards the negative pole under the 10g/L sodium dodecyl sulfate (SDS) solution. Not only are the lights generated at the moment the liquid metal contacts the copper (Figure 1b), they exist at the gap of liquid metal stream as well (Figure 1c). Figure 1d shows another case triggering the light, which is operated under the oil solution and the negative pole is some liquid metal droplets.

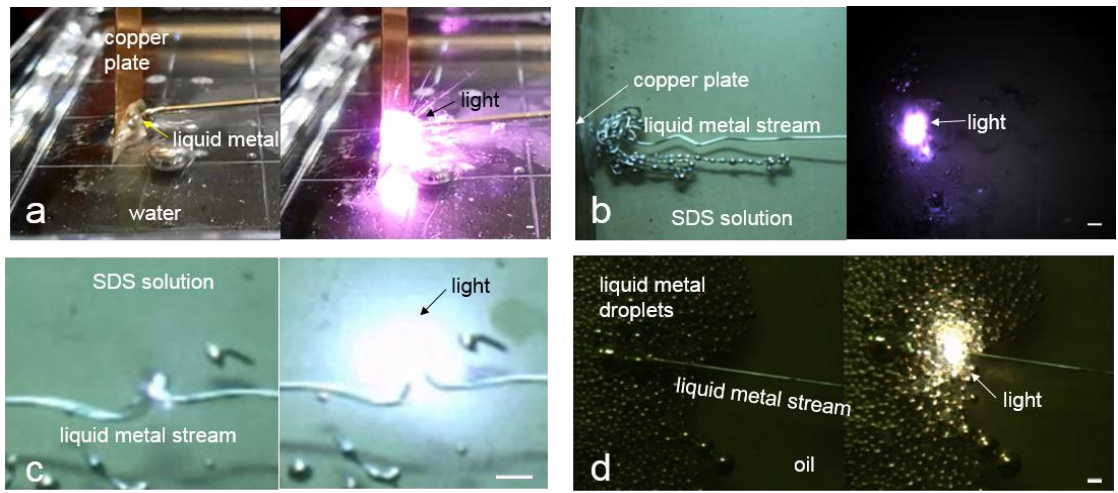

**Figure 1.** Several cases of electrically-induced light based on liquid metal. (a) Static liquid metal serves as positive electrode and copper plate as negative one, and the solution is water. Scale bar is 1mm; (b) Positive electrode is a liquid metal stream jetted to the negative electrode, and the stream is in 10g/L SDS solution. Scale bar is 1mm; (c) A gap exists during the liquid metal jetting period, and the stream is in 10g/L SDS solution. Scale bar is 1mm; (d) A liquid metal stream as the positive electrode is jetted to the liquid metal droplets connected with the negative pole, and the stream is in the oil. Scale bar is 1mm.

Besides, heat is sent off together with the emitted light. However, the thermal energy will quickly transmit to the surrounding solution quickly. Thus, only a slight temperature rise can be observed after a single burst. Also, this phenomenon is accompanied with chemical reaction generating a mass of bubble. It is noticed that the light is often generated at the gap between two metal electrodes. Specially, for all-soft liquid metal electrodes, this sandwich structure and the phenomenon could be described use the model presented in Figure 2.

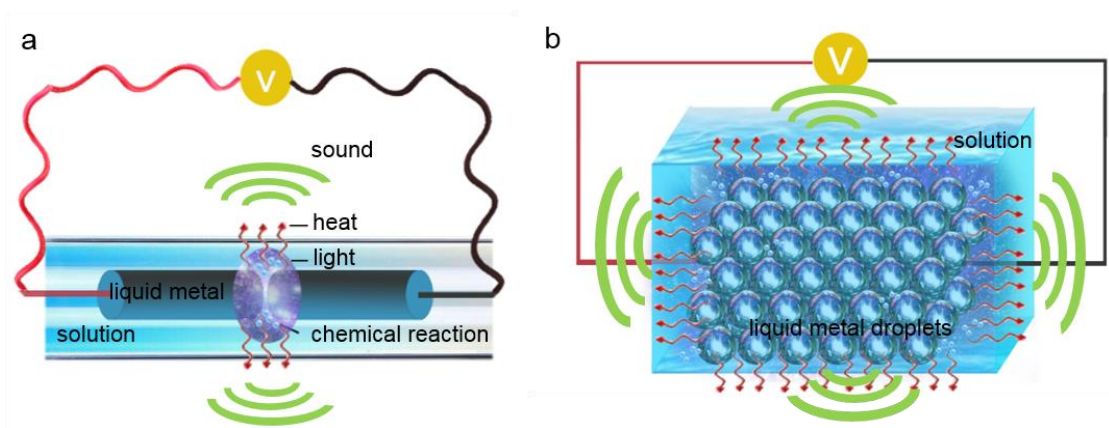

**Figure 2.** Model of electrically-induced light and sound based on liquid metal.
(a) One dimensional case; (b) Volumetric case.

In order to find out the wavelength distribution of the light, the optical emission spectrum was scanned with a fiber optic spectrometer (Ocean Optics USB2000, US). Figure 3a gives the optical spectrum of the spark emitted by the liquid metal electrodes shown in Figure 1a and Figure 1b, respectively. The emitted lights are purple visually. Nine obvious peaks (280nm-450nm and about 640nm) can be observed along the axis, which are mainly originated from the gallium, indium and water. Among them, four peaks were located in the ultraviolet area. The strongest four peaks lie in the section of visible blue and violet.

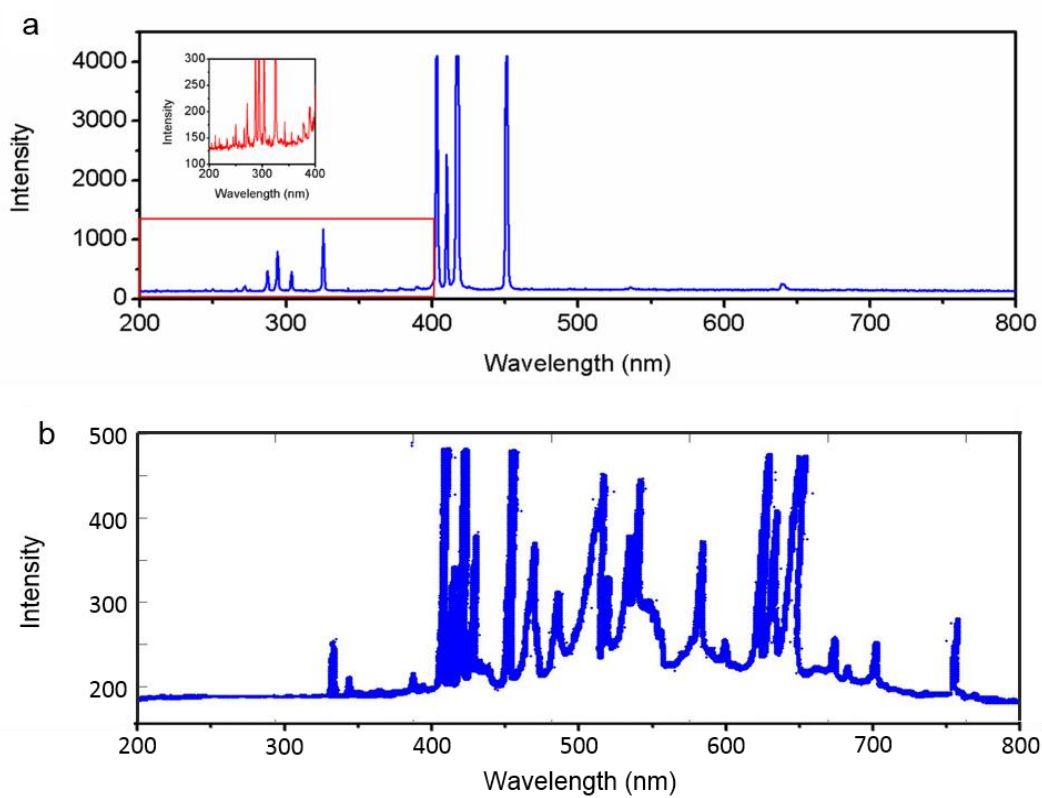

**Figure 3.** Spectrum distribution of emitted light. (a) Case of liquid metal electrodes under the SDS solution; (b) Case of liquid metal electrodes under the oil solution.

Besides, the optical spectrum of the light induced by the liquid metal electrodes under oil was also measured and shown in Figure 3b. It can be figured out that the spectrum also includes the featured lines in that of the water solution, yet extra lines with rather high intensity are obvious, especially in the region of more than 500nm (yellow and red). These extra spectrum peaks are close to the optical color of the oil, which might be from the substances in the oil. All these experiments indicate that, potential options for the solution can be many which would offer plenty of opportunity to develop future quantum generator. And, simultaneous electrophonic quantum can possibly be made to entangle together since the solution environment integrate all the quantum generation elements and energy sources together. But further quantification and measurement on such entanglement effects need another completely different work in the coming time.

## 3. Quantum Physical Mechanism of Light and Sound Generation in Solution

The electric-field internsity increases with the drawing near of the positive electrode. Suddenly, when the distance falls small to a certain extent, a high puncture current would be produced between both the liquid metal electrodes, accompanied by a light emission, heat and a sound passing through the solution, and some chemical reactions are also brought about in the solution. According to the law of conservation of energy, the processes of quantum energy transfer could be characterized as follows:

$$E_e = E_o + E_h + E_a + E_c \quad (1)$$

where, $E$ represents the energy, the subscripts of e, o, h, a and c represent electrical, optical, heat, acoustic and chemical energy, respectively.

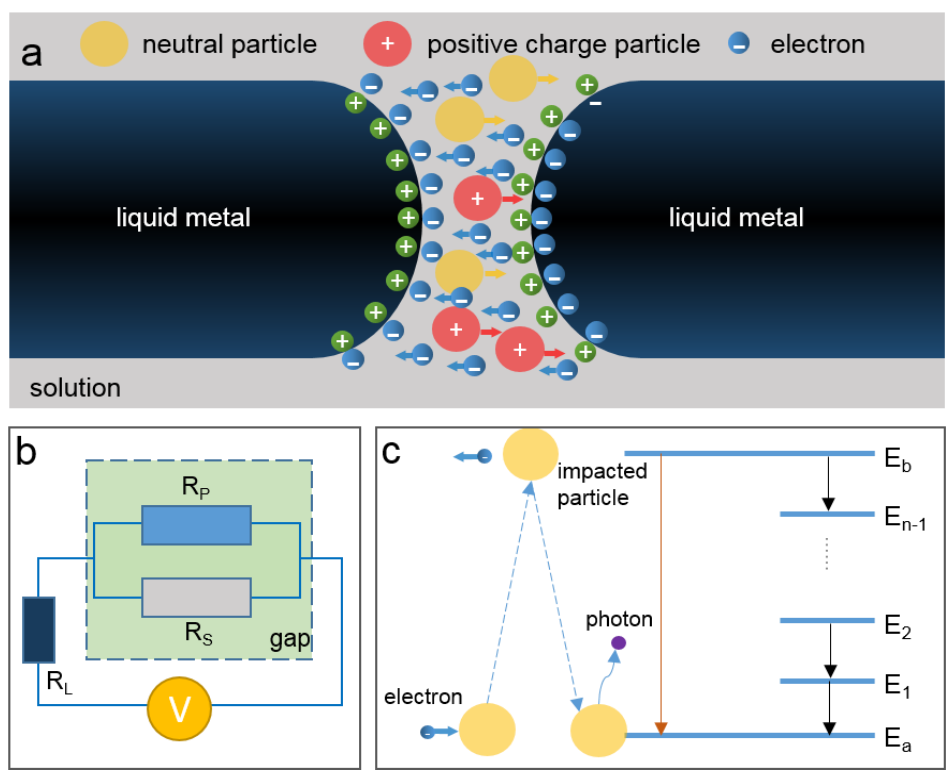

**Figure 4.** Schematic diagram of electrically-induced photonic quantum. (a) Collision among the particles distributed in the liquid metal gap; (b) Sketch of energy conversion after hitting by an electron; (c) Energy level transition of excited particle.

As shown in Figure 4a, at this gap, there exists various particles, such as electrons, moleculars (solution), atoms (gallium, indium), ions (eg. $Ga^{3+}$, $In^{3+}$, $OH^-$, $H^+$ etc.). Only the charged particles could be accelerated and gain momentum. Since the free path of electrons is much shorter, the collisons between electrons and other particles play significant roles on the energy transfer at this narrow gap. Here, we mainly consider the momentum of electrons achieved from the electric field. For an electron, the electric energy could be expressed by

$$E_e = eU_g \tag{2}$$

where, $e$ is the charge of an electron, $U_g$ is the voltage of the gap between two liquid metal poles.

Figure 4b depicts an equivalent circuit of the model in Figure 4a, and one can write out such a relation, i.e.

$$U_g = \frac{R_g}{R_l + R_g} U = U - I_t R_l \tag{3}$$

$$R_g = \frac{R_s R_p}{R_s + R_p} \tag{4}$$

where, $R_l$, $R_g$, $R_s$ and $R_p$ represent the resistance of the liquid metal, whole gap, solution between the gap and charged particles, respectively. $I_t$ is the puncture current, and $U$ is the applied voltage.

Before the gap is triggered, $R_g = R_s$. And during switching on, one has

$$R_l = \rho \frac{l}{S} = \frac{4\rho l}{\pi d^2} = \frac{4\rho l}{A^2 \pi d_{needle}^2} \tag{5}$$

where, $\rho$, $l$ are the electrical resistivity and length of the liquid metal stream, $d_{needle}$ is the inner diameter of the needle orifice, $A$ is the expansion coefficient of the liquid metal stream after injected from the needle orifice. According to the data in the previous study,[19] taking $\rho = 1/(3\times10^6)\Omega m$, $l = 0.1m$, A=2 and $d_{needle} = 0.21mm$ in the equation (4), one has $R_l = 24\mu\Omega$. Then plug $U = 20V$, $I_t \approx 20A$ into equation (2), $U_g$ could be estimated as $U_g \approx U = 20V$.

Then electrons with momentum deliver enery to other particles through a collsion process. As shown in Figure 4c, the particles including water molecules, gallium atoms, indium atoms, $Ga^{3+}$, $In^{3+}$, $OH^-$, $H^+$ would jump to a higher energy state $E_b$ from their original lower energy state $E_a$ after absorbing energy. However, the excited state is unstable, therefore these particles transit to lower energy states immediately and emit lights. If jumping to $E_a$ directly, a light is emitted, and its wavelength $\lambda$ could be calculated by

$$\lambda = \frac{c}{\nu_o} = \frac{ch}{E_b - E_a} \tag{6}$$

where, $\nu_o$ is the frequency of emitted light, $h$ is the Planck constant, $c$ is the velocity of light.

If juming to $E_a$ through the middle enegy levels, the emitting lights include several wavelength, i.e.

$$\lambda_1 = \frac{ch}{E_1 - E_a} \tag{7}$$

$$\lambda_2 = \frac{ch}{E_2 - E_1} \tag{8}$$

……

$$\lambda_n = \frac{ch}{E_b - E_{n-1}} \tag{9}$$

After collision, the electron will continue to move with the new velocity, and start the next collision and energy transfer. From the above equations, it could be found that the maximum enegy of the emited light depends on the electron momentum and the difference between the two enegy states $\Delta E = E_b - E_a$. The former is determined by the electric field intensity of the gap. And the latter depends on the excitation potentials of the particles that are hit by the electrons.

The heat energy here is caused by the random movement of a large number of particles.[20] An electron with a certain speed hits some particles, allowing them to obtain the kinetic energy. Then these particles continually hit some other particles. Thus, all the particles in the gap have obtained the non-regular kinetic energy, which is embodied at the macro level that lead to the temperature rises. Besides, the particles in the gap will also hit the particles in the solution, so that the motion of particles in the solution intensifies. On a macro level, the heat is transported through the solution. Therefore, the kinetic energy of all the participating particles could be represented on behalf of the heat energy.

In the standpoint of quantum, the lattice vibrations in metals form sound waves, which is a kind of mechanical wave, consisting of many phonons with different vibrational frequencies. Similar to photons, phonons have wave-particle duality, too. [20, 21] Thus, the energy of a phonon could be quantized as

$$E_a = h\nu_a \tag{10}$$

When the atoms are rearranged in a reactive compound to produce new compounds, the chemical energy changes, the nature of which is the breaking of chemical bonds in the reactants and the formation of chemical bonds in the product molecules. In this process, the chemical reaction involved is

$$H_2O(l) = H_2(g) + 1/2 O_2(g) \tag{11}$$

First, the liquid water change to a gas phase through the increase of internal energy. Second, two H-O bond breaks, then a H-H bond and an O-O bond forms. Thus, the chemical energy for one water molecule is expressed as

$$E_c = E_{Liquid-gas} + 2E_{H-O} - E_{H-H} - \frac{1}{2}E_{O=O} \tag{12}$$

When the pressure is $1.01 \times 10^5$ Pa and the initial temperature is 25℃, $E_{Liquid-gas}$, $E_{H-O}$, $E_{H-H}$ and $E_{O=O}$ could be calculated as

$E_{Liquid-gas} = (40800 + 4.2 \times 18(100-25))/(6.02 \times 10^{23})\,J = 0.77 \times 10^{-19}\,J$,

$E_{H-O} = 464000/(6.02 \times 10^{23})\,J = 7.71 \times 10^{-19}\,J$,

$E_{H-H} = 436000/(6.02 \times 10^{23})\,J = 7.24 \times 10^{-19}\,J$,

$E_{O=O} = 498000/(6.02 \times 10^{23})\,J = 8.27 \times 10^{-19}\,J$

Then, one can get $E_c = 4.81 \times 10^{-19}\,J$

In this case, we could then obtain the generalized equation to quantify the electrically induced photonic and acoustic quantum effect as:

$$\overset{electric}{n_e eU} = \overset{optical}{h\sum_{i=1}^{n_o} v_{oi}} + \overset{heat}{\frac{1}{2}\sum_{i=1}^{n_p} m_i v_i^2} + \overset{acoustic}{h\sum_{i=1}^{n_a} v_{ai}} + \overset{chemical}{n_{H_2O} E_c} \tag{13}$$

where, $n_e, n_o, n_p, n_a, n_{H_2O}$ are the number of electrons, photons, all particles that have no regular motion in the gap, phonons and water molecules participating in chemical reactions, respectively. Clearly, through controlling each of the terms in the above equation, one can design a specific experimental system and generate a group of desired quantum processes which can span from a wide variety of physical properties and behaviors. Further works either theoretically or experimentally can be performed along this direction.

**4. Discussion**

The three conditions mentioned above all involve liquid solution environment. In fact, this phenomenon could also be repeated in the air. As shown in Figure 5, the positive electrode is an indium block and the negative one is a graphite rod. When the two electrodes nearly come into contact with each other, a blue-violet light generates. Compared with the cases immersed in the solution, only particles including electrons, indium atoms, $In^{3+}$ ions participate this process. And there is no chemical reaction with water. Clearly, another new quantum equation similar to Eq.(13) can also be derived, which will not be given here for brief.

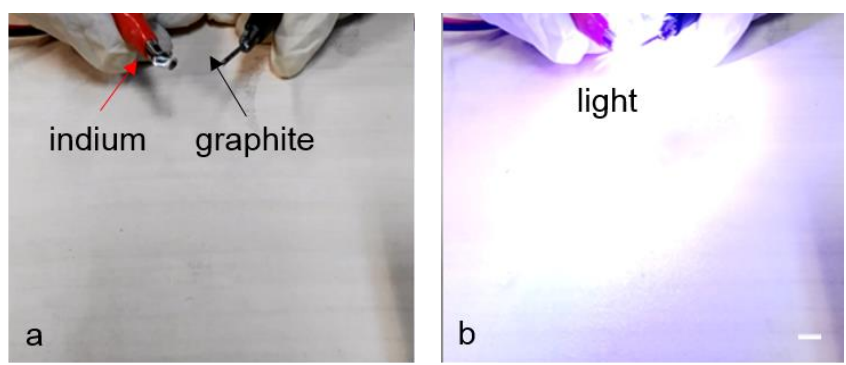

**Figure 5.** The light triggered by the indium electrode in the air. (a) Before triggering; (b) During switching on. Scale bar is 1cm.

In addition, motivated under the same external voltage, here it is 20V. There is no blue-violet light if the positive electrode is replaced with copper. As shown in Figure 6a, a solid copper wire electrode of 0.3mm diameter protected by a glass capillary tube is used. A whilte light appears when the tip of the copper wire just contacts the copper plate. And more heat energy is released because of the short circuit. Figure 6b gives the optical spectrum of the white light induced by the copper wire electrode. The results indicate that the spectrum is rather continuous which implies the case of incandescence. Comparing Figure 6c and Figure 6d, the copper wire is blackened due to the effect of heat energy. That is to say, more electrical energy has been tranfered into heat rather than light. Generally, the melting point is increased with the strength of the metal bond. Therefore, the metal bond strength of Cu-Cu is higher than Ga-Ga. $W_e$ of Cu electrode is much larger than that of Ga. This is the reason why the Ga, In or EGaIn metal electrode could emit violet light under low voltage while Cu needs extremely high voltage.

Besides, according to Figure 4b, the resistance of solution between the gap ($R_s$) and charged particles ($R_p$) are in parallel in the circuit. The greater $R_s$ is, the more energy charged particles could be obtained. This means that solution with weaker electrical conductivity generates light more easily. We chose SDS solution as a weak electrolyte and NaCl solution as a strong electrolyte. When applied with voltage, the chance of blue-violet light happening reduced for the latter and the light strength also becomes weaker. The experimental results are in good agreement with the theory, which shows that these formulas are effective.

Another potential value from the above electrophtonic quantum effect perhaps lies in that quantum entanglement may be easily induced via the present system. This is because the electricity from the same source can induce lighting and sound across multiple sites even covering the whole solution whenever many different liquid metal droplets are incorporated. The lighting quantum from different regions over the continnum solution environment can affect each other, or in orther words, entangenented among them. In this sense, if well controlled, such effects can possibly be taken full use to aid develop future quantum computing or communication system.

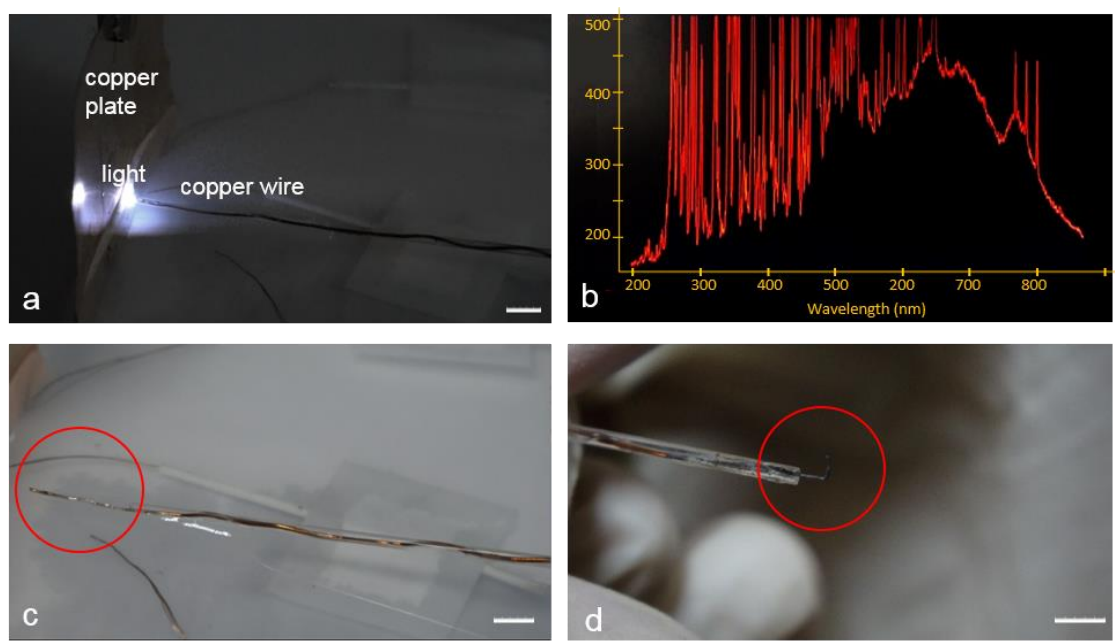

**Figure 6.** Output of copper wire as the positive electrode. (a) The light triggered by the solid copper wire electrode, Scale bar is 0.5cm; (b) Spectrum distribution of the light in (a); (c) Image of the copper wire before experiment. Scale bar is 0.5cm; (d) Image of blackened copper wire after experiment. Scale bar is 0.5cm.

5. **Conclusion**

In this article, a photonic and acoustic quantum generation system was proposed and the phenomenon thus induced by electrical field was interpreted via quantum theory through the established equation consisted of "liquid metal electrode-liquid membrane-liquid metal electrode". It is found that a small voltage applied at the ends of liquid metal structure immersed in solution could produce violet light, heat, sound and chemical reaction. Drawing on the lessons of photo-electric effect theory proposed by Einstein, we attributed such quantum reasoning as energy transformation from electric energy to light, heat, acoustic and chemical energy. This theoretical strategy will promote the development of future potential quantum computing and communication system enabled by liquid metal immersed in aqueous environment and under room temperature condition.

**Acknowledgment**

This work is partially supported by the National Natural Science Foundation of China (NSFC)

Key Project under Grant No. 91748206, Dean's Research Funding of the Chinese Academy of Sciences and the Frontier Project of the Chinese Academy of Sciences.